\documentclass[twocolumn,prl]{revtex4}%
\usepackage{amsfonts}
\usepackage[T1]{fontenc}
\usepackage{amsmath,amsbsy,amssymb,graphicx}
\usepackage{amsmath}
\usepackage{amssymb}
\usepackage{graphicx}%

\begin{document}
\title{{\Large Current Modulator based on Topological Insulator}\\{\Large with Sliding Magnetic Superlattice }}
\author{Motohiko Ezawa$^{1}$ and Jiadong Zang$^{1,2}$}
\affiliation{$^{1}$ Department of Applied Physics, University of Tokyo, Hongo 7-3-1,
113-8656, Japan }
\affiliation{$^{2}$ Department of physics, Fudan university, Shanghai 200433, China}

\begin{abstract}
We study theoretically the surface of a topological insulator with a sliding
magnetic superlattice coated above. By analyzing time-dependent Dirac
equations, the dynamics of the zero mode is investigated. When the
superlattice's sliding velocity is smaller (larger) than the Fermi velocity of
topological insulator, the zero mode is perfectly (imperfectly) pumped. We
also propose the application of this setup, which rectifies currents or
generates pulse currents. It would provide a prototype of electronic devices
based on topological insulator.

\end{abstract}
\date{\today }
\maketitle

\textit{Introduction: } Topological insulator\cite{Kane,Bernevig,Hsieh,Today}
is a new state of matter with insulating bulk and metallic surface. The
gapless surface is protected by the topology, and persisting even in the
presence of disorder as long as time reversal symmetry is respected. Both
theory and experiment have established that this surface is not a conventional
metal, but a helical liquid. As a result, a dissipationless spin current can
be realized on the surface.

When time reversal symmetry is broken on the surface, even richer properties
of topological insulator emerges, such as monopoles\cite{Zang} and chiral
Majorana fermions\cite{Majorana}. When a magnetic domain wall is attached on
the surface, a chiral mode is generated inside the wall, which shares the same
physics as Jackiw and Rebbi's zero mode\cite{Jackiw}. In this sense,
topological insulator bridges between condensed matter physics and high energy
physics. Many unrealized phenomenon in high energy physics are promisingly
addressed experimentally in topological insulator.

However, potential applications of topological insulator are seldom mentioned.
Let us consider a magnetic domain-wall lattice created on its surface. Then,
domain walls generate series of chiral zero modes, which are treated as
quantum wires. By the motion of these domains, we can tune the electronic
signals. It would provide a prototype of application in electronics.

In this paper, we find a dissipationless transverse current induced by a
magnetic superlattice attached onto a topological insulator by solving
time-dependent Dirac equations. The behavior changes drastically whether the
sliding velocity of the superlattice is larger or smaller than the velocity of
the Dirac particle. The transverse current is perfectly pumped when we apply a
slowly moving magnetic lattice (adiabatic quantum pumping) and imperfectly
pumped when we apply a fast moving lattice (non-adiabatic quantum pumping). We
also suggest an application of this set up to a current modulator, which
converts between direct, alternating, and repulse currents.

\begin{figure}[t]
\centerline{\includegraphics[width=0.27\textwidth]{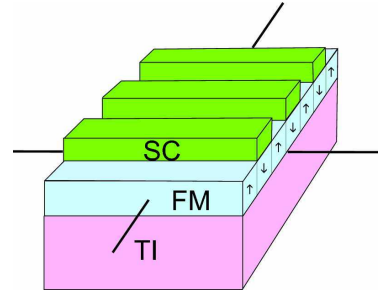}}\caption{Illustration
of an experimental setup of current modulator on the surface of a topological
intulator. }%
\label{FigSetup3D}%
\end{figure}

\textit{Model: }Our system consists of a three-dimensional topological
insulator with a ferromagnet attached onto it [Fig.\ref{FigSetup3D}].
Electrons obey 2D Dirac equations on the surface of a topological insulator,
as has been demonstrated by the spin- and angle-resolved photoemission
spectroscopy\cite{HsiehN}. In the presence of a ferromagnetic layer on top of
a topological insulator, the Hamiltonian is given by%
\begin{equation}
H=\hbar v_{\text{F}}(k_{x}\sigma_{x}+k_{y}\sigma_{y})+\mu_{B}\mathbf{M}\left(
\mathbf{x},t\right)  \cdot\mathbf{\sigma}, \label{BasicHamil}%
\end{equation}
where $v_{\text{F}}$ is the Fermi velocity of Dirac fermions, which is
$3\times10^{5}$m/s for Bi$_{2}$Se$_{3}$\cite{Zhang}, $\sigma_{i}$ are the
Pauli matrices, $\mu_{B}$ is the Bohr magneton, and $\mathbf{M}\left(
\mathbf{x},t\right)  $ represents the strength of the exchange coupling. It
has been argued\cite{Yokoyama} that the in-plane magnetic field contributes
only tiny effects to the transport property of a topological insulator.
Namely, it is a good approximation that $\mathbf{M}\left(  \mathbf{x}%
,t\right)  $ has only the $z$-component, $\mathbf{M}\left(  \mathbf{x}%
,t\right)  =\left(  0,0,M\left(  x,t\right)  \right)  $. Let us assume that
the system is homogeneous along the $y$-axis.\ Then, the central point is the
$x$ and $t$ dependence of this exchange coupling, which is realized by domain
wall motion along the $x$-axis with velocity $v$ in the ferromagnet layer. It
is reasonable to set
\begin{equation}
M\left(  x,t\right)  =m\sin\left[  \left(  x-vt\right)  /\lambda\right]
\label{DWsin}%
\end{equation}
to describe a domain-wall lattice, where $m$ is a numerical constant
characterizing the strength of a single domain. We define $\omega=v/\lambda$.

\textit{Static solution: }The sinusoidal behavior of $M(x,t)$ leads to a
series of zero-field points periodically. As particle-hole symmetry is
respected, according to Jackiw and Rebbi's work, zero modes appear around
these points. In order to derive the explicit expression of the zero modes
here, let's start from the time-dependent Schr\"{o}dinger equation
\begin{equation}
i\hbar\partial_{t}\Psi=H\Psi, \label{TSchr}%
\end{equation}
where we have set the wave function as $\Psi=\left(  \psi_{\uparrow}%
,\psi_{\downarrow}\right)  ^{t}$. We may set $k_{y}=$constant in
(\ref{BasicHamil}) due to the translational invariance along the $y$-axis.

\begin{figure}[t]
\centerline{\includegraphics[width=0.45\textwidth]{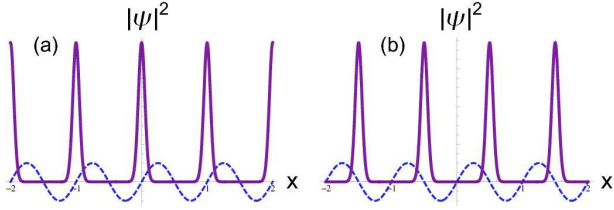}}\caption{We
illustrate the magenetization due to the magnetic superlattice $M\left(
x\right)  $ (dashed curve) and the static solutions $\left\vert \psi\left(
x\right)  \right\vert ^{2}$ associated with it (solid curve), (a) for the $+$
sign and (b) for the $-$ sign in (\ref{TypeSine}). The horizontal line
represents the $x$ axis. We have set $\mu_{B}m\lambda/(\hbar v_{\text{F}}%
)=5$.}%
\label{FigStatic}%
\end{figure}

We first study the static case where $\omega=0$. Particle-hole symmetry
guarantees the existence of zero-energy solutions with the relation
$\psi\equiv\psi_{\uparrow}=\pm i\psi_{\downarrow}$. When $k_{y}=0$, the
equation of motion is transformed into
\begin{equation}
\mu_{B}M\left(  x\right)  \psi=\pm\hbar v_{\text{F}}\partial_{x}\psi.
\end{equation}
We may solve it as\cite{Jackiw}
\begin{equation}
\psi_{\pm}\left(  x\right)  =\exp\left[  \pm\frac{\mu_{B}}{\hbar v_{\text{F}}%
}\int_{0}^{x}M\left(  x^{\prime}\right)  dx^{\prime}\right]  ,
\label{JackiRebbi}%
\end{equation}
which yields
\begin{equation}
\psi_{\pm}\left(  x\right)  =\exp\left[  \pm\frac{\mu_{B}m\lambda}{\hbar
v_{\text{F}}}\cos\frac{x}{\lambda}\right]  , \label{TypeSine}%
\end{equation}
up to a normalization constant. We illustrate the magnetization and
$\left\vert \psi\left(  x\right)  \right\vert ^{2}$ for these two cases in
Fig.\ref{FigStatic}.

We may also present solutions for the non-zero $k$ case. The two solutions
have opposite chirality, so their group velocity along $y$-direction is
opposite as well. The wave function is a linear combination of $\psi_{\pm
}\left(  x\right)  $,
\begin{equation}
\Psi=e^{ik_{y}y}\Psi_{+}+e^{-ik_{y}y}\Psi_{-},
\end{equation}
with (\ref{TypeSine}).

\textit{Moving Domain Wall:} We start with the investigation of the domain
wall motion described by (\ref{DWsin}). The sinusoidal potential has a
double-periodicity,
\begin{equation}
M\left(  x+L,t\right)  =M\left(  x,t\right)  ,\quad M\left(  x,t+T\right)
=M\left(  x,t\right)  ,
\end{equation}
where $L=2\pi\lambda$ and $T=2\pi\lambda/v$. To such a system both the Bloch
theorem and the Floquet theorem are applicable: The wave function is of the
form $\psi_{\alpha}\left(  x,t\right)  =e^{i\varepsilon_{\alpha}t+ik_{\alpha
}x}u_{\alpha}\left(  x,t\right)  $ with $u_{\alpha}\left(  x+L,t\right)
=u_{\alpha}\left(  x,t\right)  $ and $u_{\alpha}\left(  x,t\right)
=u_{\alpha}\left(  x,t+T\right)  $. Then it is enough to analyze the torus
region $0\leq x<L,0\leq t<T$.

\begin{figure}[t]
\centerline{\includegraphics[width=0.44\textwidth]{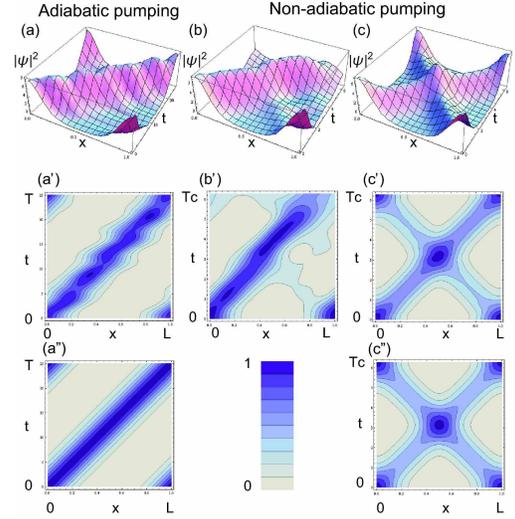} \label{FigMag}}\caption{We
show numerical solutions of the time-dependent Dirac equation with the
sinusoidal potential for (a,a') $\omega/\omega_{\text{c}}=0.25$, (b,b')
$\omega/\omega_{\text{c}}=3$, (c,c') $\omega/\omega_{\text{c}}=10$. The
vertical axis is $\left\vert \psi\left(  x,t\right)  \right\vert ^{2}$. The
horizontal axes are $x$ and $t$. (a), (b) and (c) are bird's eye view and
(a'), (b') and (c') are their contour plot. We also present analytic solutions
in (a") and (c"), which are valid $\omega/\omega_{\text{c}}\ll1$ and
$\omega/\omega_{\text{c}}\gg1$, respectively. The figures are plotted in the
region $0<x<L,0<t<T$ for the adiabatic case, and in the region
$0<x<L,0<t<T_{\text{c}}$ for the non-adiabatic case, where $T_{\text{c}%
}=(\omega/\omega_{\text{c}})T$. We have started from the static solution with
$\mu_{B}m\lambda/(\hbar v_{\text{F}})=1$.}%
\label{FigSin}%
\end{figure}

First we investigate the case where the velocity of the moving lattice is low,
$v<v_{\text{F}}$, or $\omega<\omega_{\text{c}}$ with $\omega_{\text{c}%
}=v_{\text{F}}/\lambda$. We have carried out a numerical analysis of the
time-dependent Dirac equation, starting from a static solution, whose results
we give in Fig.\ref{FigSin}(a,a'), where we have set $\omega/\omega_{\text{c}%
}=0.25$. We find that the zero mode moves together with the magnetic lattice,
and it is adiabatically pumped.

It is possible to derive the analytic solution based on the adiabatic
approximation,
\begin{equation}
\psi_{\pm}\left(  x,t\right)  =\exp\left[  \pm\frac{\mu_{B}m\lambda}{\hbar
v_{\text{F}}}\cos\frac{\left(  x-vt\right)  }{\lambda}\right]  ,
\label{Galilei}%
\end{equation}
which is constructed by making the Galilei boost of the static solution
(\ref{JackiRebbi}): See Fig.\ref{FigSin}(a"). This solution is valid when the
velocity is very low. There are fluctuations in $\left\vert \psi\left(
x,t\right)  \right\vert $ in the case of the numerical calculation, which is
absent in the analytic one. This is because the adiabatic solution
(\ref{Galilei}) is not an exact solution of (\ref{TSchr}). In any case, if we
introduce the function $P_{n}(t)$ representing the position of $\left\vert
\psi\left(  x,t\right)  \right\vert $'s n-th peak at time $t$ in unit of
$\lambda$, we have
\begin{equation}
P_{n}\left(  T\right)  -P_{n}\left(  0\right)  =1
\end{equation}
for $\omega<\omega_{\text{c}}$. In that sense, the zero mode is adiabatically
pumped along $x$-direction as long as\ the velocity of domain wall motion is
smaller than the Fermi velocity. As the Fermi velocity for topological
insulator is generally very large, this requirement is usually satisfied. Due
to the periodicity along $x$-direction, $P_{n}(t)$ actually doesn't depend on
$n\,$, and the subscript is neglected in the following.

It's also interesting to investigate the case where the velocity of the moving
lattice is fast, $v>v_{\text{F}}$, or $\omega>\omega_{\text{c}}$. We have
carried out a numerical analysis, whose results we give in Fig.\ref{FigSin}%
(b,b') and (c,c'), where we have set $\omega/\omega_{\text{c}}=3$ and $10$.
Fig.\ref{FigSin}(b,b') are plotted in the region $0<t<T$, while
Fig.\ref{FigSin}(c,c') are plotted in the region $0<t<T_{\text{c}}$ with
$T_{\text{c}}=(\omega/\omega_{\text{c}})T$.\ 

In the large $\omega$ limit, we give also an analytic solution. Since the
magnetization oscillates very quickly, we can approximate the system by the
one-period time-averaged equation of motion\cite{MCloskey},
\begin{equation}
i\hbar\partial_{t}\psi=\overline{H}\psi,
\end{equation}
where
\begin{equation}
\overline{H}=\frac{1}{T}\int_{0}^{T}H\left(  x,t\right)  dt.
\end{equation}
The equation of motion is rewritten as
\begin{equation}
i\hbar\partial_{t}\psi=-i\hbar v_{\text{F}}\left(  \partial_{x}\sigma
_{x}+\partial_{y}\sigma_{y}\right)  \psi,
\end{equation}
or
\begin{equation}
\partial_{t}^{2}\psi=v_{\text{F}}^{2}\left(  \partial_{x}^{2}+\partial_{y}%
^{2}\right)  \psi.
\end{equation}
It does not depend on $\omega$. The solution is given by $\left[  \psi\left(
x-v_{\text{F}}t\right)  +\psi\left(  x+v_{\text{F}}t\right)  \right]  /2$ with
(\ref{TypeSine}), which we display in Fig.\ref{FigSin}(c"). The agreement is
good between the analytic and numerical solutions when the velocity is large
($\omega/\omega_{\text{c}}>10$).

The pumping rate is independent of $\omega$. It implies that one transverse
current is pumped in the period which is not $T$ but $T_{\text{c}}$,
\begin{equation}
P\left(  T_{\text{c}}\right)  -P\left(  0\right)  =1.
\end{equation}
The primitive torus $0<t<T$ has been extended to $0<t<T_{\text{c}}$. As a
result, the pumping transverse current in one period $T$ is anti-proportional
to the frequency,
\begin{equation}
P\left(  T\right)  -P\left(  0\right)  =\omega_{\text{c}}/\omega.
\end{equation}
The reason of this imperfect pumping is that the transverse current can not
follow the magnetic lattice because it moves too fast.

\begin{figure}[t]
\centerline{\includegraphics[width=0.37\textwidth]{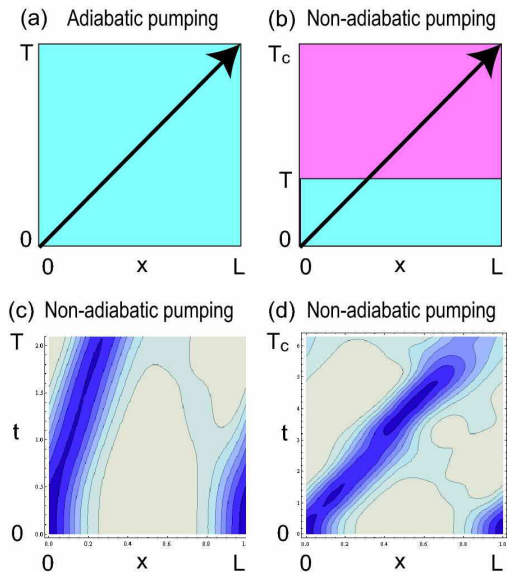}}\caption{Schematic
diagram of the adiabatic (a) and non-adiabatic (b) pumping. One transverse
current is pumped in the time $T$ in the adiabatic pumping, while one
transverse current is pumped in the time $T_{\text{c}}$ in the non-adiabatic
pumping. Numerical results for the non-adiabatic pumping for the period (c)
$[0,T]$, and (d) $[0,T_{\text{c} }]$.}%
\label{FigIllustPump}%
\end{figure}

A comment is in order. The adiabatic pumping case is experimentally realsitic,
since it is hard to move domain walls faster than $v>v_{\text{F}}$
experimentally at this stage. Nevertheless the non-adiabatic case is
theoritically very interesting, as we have shown.

\begin{figure}[t]
\centerline{\includegraphics[width=0.34\textwidth]{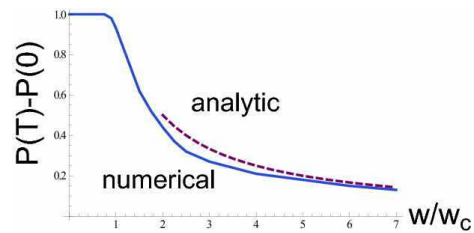}}\caption{$\omega$
dependence of $P\left(  T\right)  -P\left(  0\right)  $. The horizontal axis
is $\omega$ and the vertical axis is $P\left(  T\right)  -P\left(  0\right)
.$}%
\label{PumpCharge}%
\end{figure}

In order to analyze the transition between the adiabatic and non-adiabatic
pumpings, we calculate the pumping transverse current numerically. The pumping
rate is estimated from the velocity of the position where $\left\vert
\psi_{\uparrow}\left(  x,t\right)  \right\vert $ takes the maximum value at
$t=T$. We show the numerical solution in Fig.\ref{PumpCharge}. It is seen that
the pumping transverse current is well described by
\begin{equation}
P\left(  T\right)  -P\left(  0\right)  =\left\{
\begin{array}
[c]{ccc}%
1 & \text{for} & \omega<\omega_{\text{c}}\\
\omega_{\text{c}}/\omega & \text{for} & \omega\gg\omega_{\text{c}}%
\end{array}
\right.  .
\end{equation}

The analytic and numerical solutions show a good agreement. The
change between the adiabatic and non-adiabatic transverse current pumping
takes place rather suddenly. It may be a kind of a non-equiribrium phase transition.

\textit{Experimental setup: }The realization of magnetic superlattice is a
highly nontrivial question. Here we present an experimental proposal designed
in the adiabatic pumping regime, as illustrated in Fig.\ref{FigSetup3D}. On
top of a topological insulator, a ferromagnetic thin layer is coated. Above
the ferromagnetic layer, an array of periodically distributed superconductors
is deposited. Usually there are many domains in the ferromagnet, which are
naturally generated and hard to control. However, with the help of this array
of superconductors, it becomes possible. At the beginning, an upward external
magnetic field magnetizes the whole ferromagnetic layer in the same direction.
Then, we lower the temperature below the critical temperature of
superconductor, and reverse the direction of external magnetic field. Due to
the Meissner effect, magnetic field is screened just below the superconductor,
and unscreened elsewhere. As a result, magnetization in the areas without
superconductor above is reversed, and the magnetization below the
superconductor may remain unchanged if the external magnetic field is properly
controlled. We would obtain a magnetic superlattice on top of a topological
insulator in this way.

It is well known\cite{Yamaguchi} that a current can drive the magnetic motion
of domain walls. By introducing a current in the ferromagnetic layer along the
$x$-direction, we can approximately realize the moving magnetic configuration
in (\ref{DWsin}), and conducting channels provided by the zero modes are
generated. Once the device is bridged by a conducting channel, a significant
current is detected.

It's an interesting property that the two zero modes in one period of the
magnetic superlattice have opposite chiralities. These two conducting channels
have tendency to transport opposite currents. We propose three type devices.

(a) When we do not apply the current into the ferromagnet, the magnetic
superlattice is static. We apply an alternating voltage between two leads
parallel to the domain wall. The resulting current is rectified into a direct
current,%
\begin{equation}
J_{\text{out}}=\max\left[  J_{\text{in}},0\right]  ,
\end{equation}
because the zero-energy conducting channel is chiral and hence only the
forward bias current goes through. This system rectifies alternating currents
to direct currents, or acts as a diode.

\begin{figure}[t]
\centerline{\includegraphics[width=0.48\textwidth]{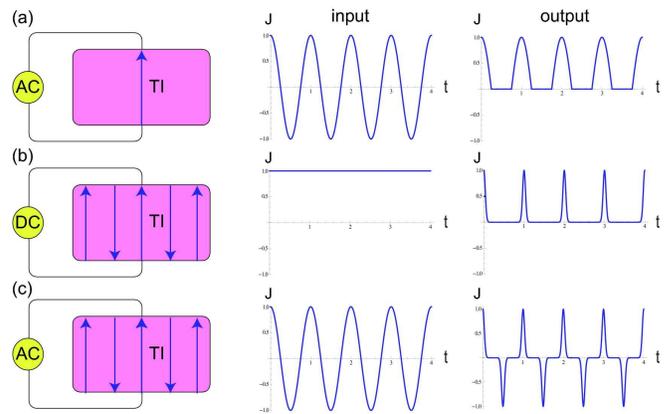}}\caption{We show the
time dependence of transverse current. (a) current rectifier (b) pulse
generator (c) alternating pulse generator. We have set $\mu_{B}m\lambda/(\hbar
v_{\text{F}})=10$.}%
\label{FigJR}%
\end{figure}

(b) When we apply the current into the ferromagnet, the domain wall
moves\cite{Yamaguchi}. We apply the direct voltage between two leads. The
resulting current is modulated to be a pulse-current,
\begin{align}
J_{+}\left(  t\right)  =  &  \operatorname{Re}\left[  \frac{\hbar}{2mi}%
\Psi_{+}^{\dagger}\partial_{y}\Psi_{+}\right] \nonumber\\
=  &  \frac{\hbar k_{y}}{2m}\exp\left[  2\frac{\mu_{B}m\lambda}{\hbar
v_{\text{F}}}\cos\frac{\left(  x-vt\right)  }{\lambda}\right]  ,
\end{align}
\newline because the zero energy conducting channel passes through the contact
periodically according to the motion of domain walls. This system acts as a
pulse generator.

(c) When we apply an alternating voltage to the same setup with (b), the
resulting signal between two devices is an alternating pulse. This current is
given by%
\begin{align}
J\left(  t\right)  =  &  \operatorname{Re}\left[  \frac{\hbar}{2mi}%
\Psi^{\dagger}\partial_{y}\Psi\right] \nonumber\\
=  &  \frac{\hbar k_{y}}{m}\sinh\left[  2\frac{\mu_{B}m\lambda}{\hbar
v_{\text{F}}}\cos\frac{\left(  x-vt\right)  }{\lambda}\right]  ,
\end{align}
and is shown in Fig.\ref{FigJR}. This system acts as an alternating pulse generator.

It will be possible to measure these currents by attaching leads to the
topological insulator parallel to the domain wall as in Fig.\ref{FigSetup3D}.

\textit{Conclusion: }In this paper, we have studied the dynamics of the zero
mode in the presence of a magnetic superlattice on top of a topological
insulator, which is described by a time-dependent Dirac equation. The domain
wall motion of the superlattice is studied.\ We have proposed a prototype of
electronic device based on the this theoretical studies, where the input
current is found to be significantly modulated by the magnetic superlattice
[Fig.\ref{FigSetup3D}]. We hope this work can promote the application of
topological insulator in future.

This work was supported in part by Grants-in-Aid for Scientific Research from
the Ministry of Education, Science, Sports and Culture No. 20940011.

\end{document}